\documentclass[sn-nature]{sn-jnl}

\usepackage{graphicx}%
\usepackage{multirow}%
\usepackage{amsmath,amssymb,amsfonts}%
\usepackage{amsthm}%
\usepackage{mathrsfs}%
\usepackage[title]{appendix}%
\usepackage{xcolor}%
\usepackage{textcomp}%
\usepackage{manyfoot}%
\usepackage{booktabs}%
\usepackage{algorithm}%
\usepackage{algorithmicx}%
\usepackage{algpseudocode}%
\usepackage{listings}%

\usepackage[superscript,biblabel]{cite}

\theoremstyle{thmstyleone}%
\theoremstyle{thmstyletwo}%

\theoremstyle{thmstylethree}%

\begin{document}

\title[Article Title]{Magnetic-field periodic quantum Sondheimer oscillations in thin-film graphite}


\author*[1]{\fnm{Toshihiro} \sur{Taen}}\email{taen@issp.u-tokyo.ac.jp}

\author[1]{\fnm{Andhika} \sur{Kiswandhi}}\email{kiswandhi@issp.u-tokyo.ac.jp}

\author[1]{\fnm{Toshihito} \sur{Osada}}\email{osada@issp.u-tokyo.ac.jp}

\affil*[1]{\orgdiv{The Institute for Solid State Physics}, \orgname{The University of Tokyo}, \orgaddress{\street{5-1-5 Kashiwanoha}, \city{Kashiwa}, \postcode{277-8581}, \state{Chiba}, \country{Japan}}}

\abstract{Materials with the mesoscopic scales have provided an excellent platform for quantum-mechanical studies. Among them, the periodic oscillations of the electrical resistivity against the direct and the inverse of the magnetic fields, such as the Aharonov-Bohm effect \cite{PhysRev.115.485,PhysRev.123.1511,PhysRevLett.5.3,PhysRevLett.56.792} and the Shubnikov-de Haas effect\cite{PhysRevB.83.035122,PhysRevLett.108.117401,shoenberg_1984}, manifest the interference of the wavefunction relevant to the electron motion perpendicular to the magnetic field. In contrast, the electron motion along the magnetic field also leads to the magnetic-field periodicity, which is the so-called Sondheimer effect\cite{PhysRev.80.401}. However, the Sondheimer effect has been understood only in the framework of the semiclassical picture\cite{JETP.8.464,PhysRev.172.718}, and thereby its interpretation at the quasiquantum limit was not clear. Here, we show that thin-film graphite exhibits clear sinusoidal oscillations with a period of about 1-3 T over a wide range of the magnetic fields (from around 10 T to 30 T), where conventional quantum oscillations are absent\cite{PhysRevB.98.155136}. In addition, the sample with a designed step in the middle for eliminating the stacking disorder effect verifies that the period of the oscillations is inversely proportional to the thickness, which supports the emergence of the Sondheimer oscillations in the quasiquantum limit. These findings suggest that the Sondheimer oscillations can be reinterpreted as inter-Landau-level resonances even at the field range where the semiclassical picture fails. Our results expand the quantum oscillation family, and pave the way for the exploration of the out-of-plane wavefunction motion.}




\maketitle

Oscillatory transport behaviour under a magnetic field is a hallmark of the quantum transport property.
For example, the Shubnikov-de Haas oscillations (SdHOs) have been explored in a wide class of semimetals or semiconductors \cite{PhysRevB.83.035122,PhysRevLett.108.117401,shoenberg_1984},
in which the electrical conductivity periodically oscillates against the inverse of the magnetic field.
In contrast to the SdHOs, the Aharonov-Bohm (AB) effect \cite{PhysRev.115.485,PhysRev.123.1511,PhysRevLett.5.3,PhysRevLett.56.792} is known as a prototypical example of the direct magnetic-field periodic oscillations \cite{Nature.397.673,PhysRevB.77.085413}.
These quantum transport features stem from the wave nature of electrons along the path perpendicular to the magnetic field through quantizations or interferences.
On the other hand, such an oscillatory behaviour associated with the electron motion along the magnetic field was absent,
since the electron motion along the magnetic field is, in principle, not influenced.
Exceptional examples were found in very clean systems with a long mean free path,
such as metallic or semiconducting crystals \cite{JPhysSocJpn.19.2353,PhysRev.172.718,PhysRevB.8.5567,PhysLettA.53.227,JLowTempPhys.36.79,PhysStatusSolidiB.94.309,JLowTempPhys.38.267}, mesoscopic-scale systems \cite{NatCommun.12.4799,NanoLett.22.65}, and the surface of a topological insulator \cite{PhysRevB.84.125144},
in which clear magnetic-field-periodic oscillations are observed.
These oscillations are known as the Sondheimer effect \cite{PhysRev.80.401},
which is determined by the 'extra' momentum of the helical motion in a finite length scale (thickness of the film),
and can be simply understood as a semiclassical effect \cite{JETP.8.464,PhysRev.172.718}.

Bernal-stacked graphite (Fig. \ref{fig1}a) provides an ideal platform in the study of the thickness-dependent features
since it has a simple layered structure of two-dimensional carbon sheets (graphene) and hence we can easily obtain a sample with a well-defined thickness without any surface reconstructions.
Moreover, the structural disorder is small enough to study the quantum transport properties.
In fact, the mobility of the graphene isolated from graphite crystals is outstanding, 
which is experimentally demonstrated in state-of-the-art devices such as electron interferometers \cite{NatPhys.5.222,NatCommun.4.2342,NatPhys.9.225,PhysRevLett.113.116601,JPhysSocJpn.84.121007,SciAdv.3.e1700600,NatNanotechnol.16.563} as well as the confirmation of the hydrodynamic flow \cite{Science.351.6277,Science.351.1058,NatPhys.13.1182,Nature.576.75,NatNanotechnol.16.563}.
Remarkably, its low carrier concentration enables us to reach the so-called quasiquantum limit at an accessible magnetic field (around 8 T perpendicular to the plane),
where only two Landau levels (LLs) remain at the Fermi level \cite{PhysRevLett.102.166403}.
In this quasiquantum limit, the motion of carriers can no longer be treated in the semiclassical picture.

Here, we perform the quantum transport measurements in thin-film graphite under the magnetic field perpendicular to the film in order to gain insight into the Sondheimer effect in the quasiquantum limit.
We observed clear sinusoidal oscillations periodic in the magnetic field in this field regime, with a thickness-dependent period, in contrast to the Shubnikov-de Haas oscillations.
We prepared thin-film graphite mechanically exfoliated from Kish graphite crystals and transferred it onto a silicon substrate to form the field-effect transistors (Fig. \ref{fig1}b,c).
The transverse magnetoresistance up to 35 T was measured, as detailed previously \cite{PhysRevB.98.155136}.
Notably, the thickness of our samples is in the range of 50 to 100 nm,
corresponding to the order of 100 unit cells,
where the quantum size effect plays an important role \cite{PhysRevB.97.115122,PhysRevB.98.155136}.
Figure \ref{fig1}d represents the magnetoresistance in the thin-film graphite with a thickness of 70 nm.
Below 8 T, the SdHO patterns are the same as that in bulk crystals \cite{PhysRevLett.102.166403},
which demonstrates that the three-dimensional band dispersion still holds.

%


\section*{Magnetic-field-periodic resistivity oscillations}\label{sec1}

\begin{figure}[h]%
\centering
\includegraphics[width=0.9\textwidth]{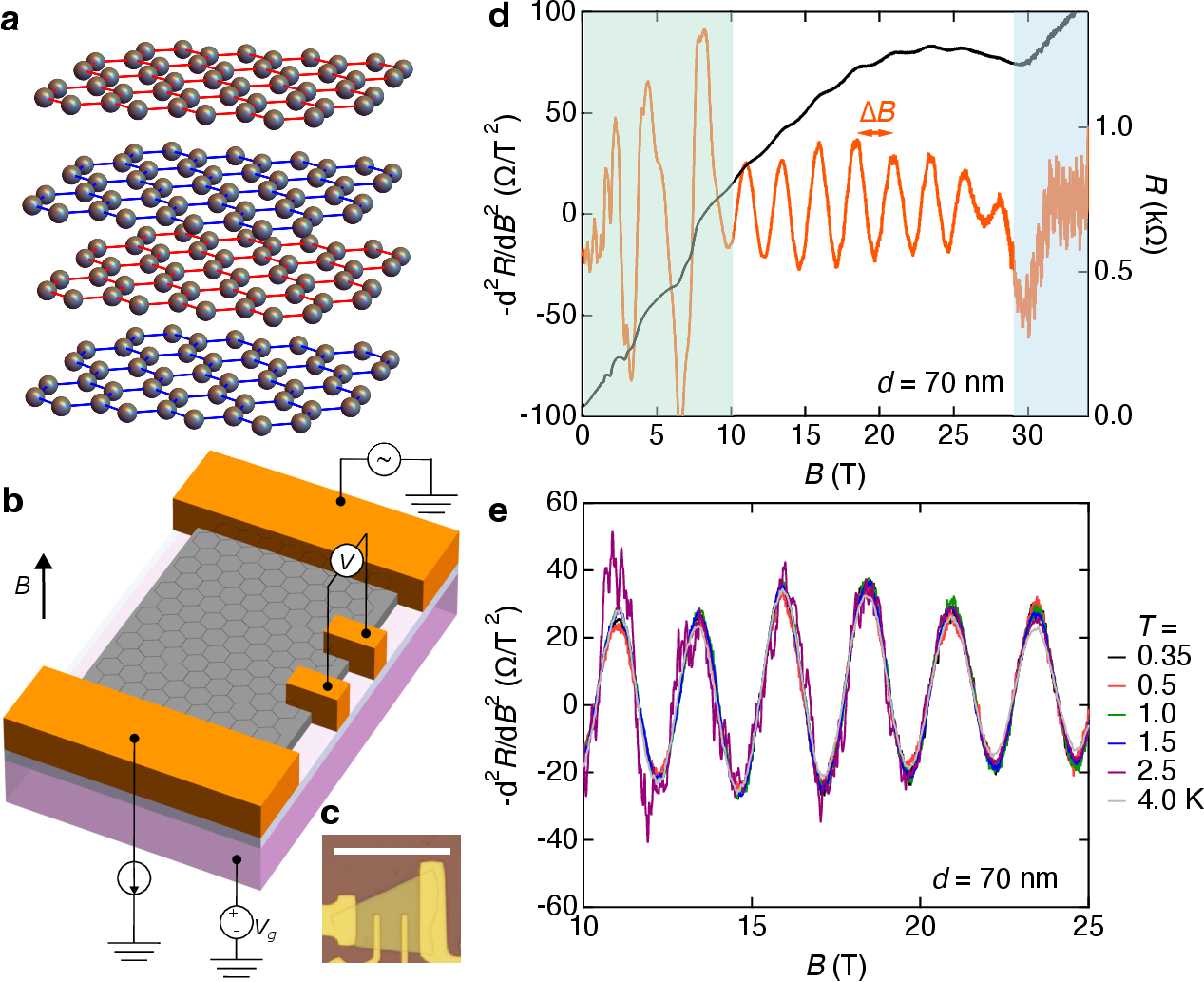}
\caption{Magnetic-field periodic resistivity oscillations. (a) Schematic crystal structure of Bernal-stacked graphite. (b) Schematic of the thin-film graphite device. Transverse resistance ($R=V/I$) was measured by sweeping the magnetic fields ($B$) applied perpendicular to the film. (c) Optical microscopy image of the device 1 ($d=70\;\mathrm{nm}$); scale bar, $50\;\mathrm{\mu m}$. (d) Magnetic field dependence of the transverse resistance $R$ and its second derivative with respect to the magnetic field. At low magnetic fields (green shaded region), the SdHOs --periodic in the inverse of the magnetic fields-- are observed. Above $30\;\mathrm{T}$, the system is in the insulating density-wave phase (blue shaded region). In between, in $10-30\;\mathrm{T}$, clear $B$-periodic resistivity oscillations observed. (e) The oscillatory components of the resistance estimated by the second derivative between 10 and 25 T at different temperatures.}\label{fig1}
\end{figure}

In a bulk semimetal or semiconductor, any anomalous transport behaviour is not expected between the SdHOs unless many-body effects play a crucial role.
In fact, once the system goes into the quasiquantum limit at around 8 T,
where the LL with an index $N=1$ escapes from the Fermi energy and no quantum oscillations occur up to 53 T \cite{JPhysCondMatt.10.11315,PhysRevLett.110.266601,JPhysSocJpn.84.054709},
the bulk graphite does not exhibit any notable structure in the magnetoresistivity before entering the electron-interaction-induced phase transition at around 30 T \cite{Tanuma_hmf,PhysicaB.256.621,JPhysCondMatt.21.344207,PhysRevLett.110.266601} (Fig. \ref{fig1}d).
However, our thin-film graphite shows a clear oscillatory behaviour with a single period $\Delta B \approx 2.5\;\mathrm{T}$ in a wide range of magnetic field from 10 T to 30 T, as shown in Fig. \ref{fig1}d.
As we measured in a different thickness sample,
the period of the oscillations has different values.
For example,
device 2 with $d=178\;\mathrm{nm}$ has a smaller period of $\Delta B \approx 1\;\mathrm{T}$ \cite{PhysRevB.98.155136}.
As shown in Fig \ref{fig1}e, the period and the amplitude, as well as the phase, are almost unchanged below 4 K.
In the case of the multilayer graphene stacking with a magic twist angle,
correlated effects, such as the superconductivity, are fragile above 4 K\cite{Nature.556.80,Nature.556.43,Science.365.605}.
This implies that the present temperature-insensitive oscillations observed in the less confined graphite samples than graphene are not attributable to the many-body effect.

\section*{Semiclassical and Quantum Sondheimer effect}\label{sec2}

\begin{figure}[h]%
\centering
\includegraphics[width=0.6\textwidth]{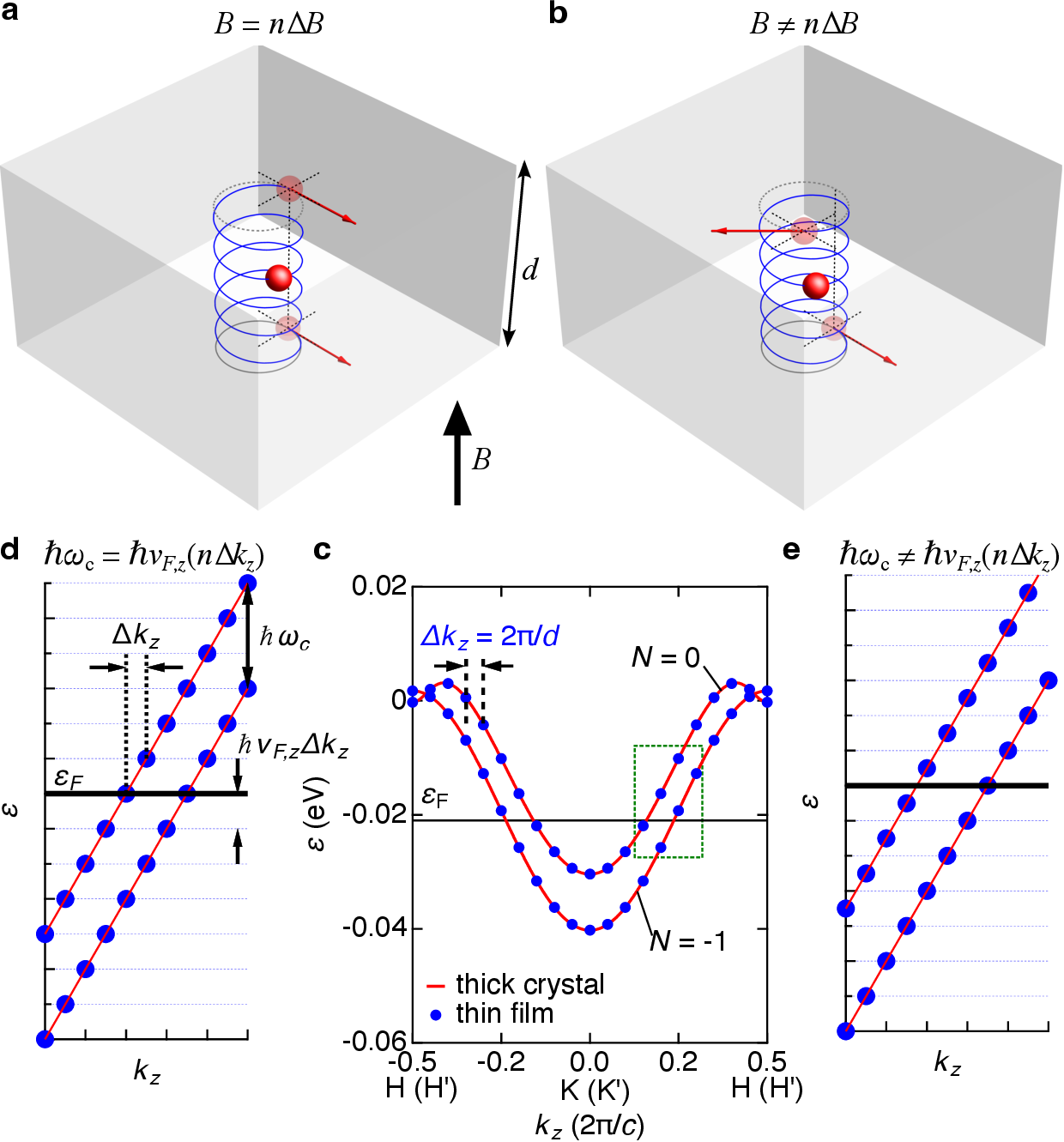}
\caption{Semiclassical and quantum pictures of the Sondheimer effect. (a,b) Schematics of the semiclassical Sondheimer effect for a clean-limit thin film with a thickness of $d$ when the magnetic field $B$ applied perpendicular to the plane is (a) equal to and (b) not equal to an integer ($n$) multiple of $\Delta B$.
Under the magnetic field, the carrier (red ball) moves along the helical orbit (blue curve). At the condition $B=n\Delta B$ (a), the integer multiple of the helical-motion period fits perfectly into the film (the black dotted cross at the top surface is located exactly above the one at the bottom). 
(c) The dispersion of LLs for graphite along $k_{z}$ at the quasiquantum limit. At the quasiquantum limit ($B>7.4\;\mathrm{T}$ perpendicular to the plane), only $N=0$ and $-1$ LLs remain on the Fermi level $\varepsilon_{F}$. In contrast to a thick system (conventional crystal), where the dispersion is continuous (red curves), $k_{z}$ in thin film is discritezed owing to the quantum size effect with a reciprocal-lattice spacing of $\Delta k_{z}=2\pi /d$ (blue solid markers). (d,e) The quantum interpretation of the Sondheimer effect. The condition Eq. (\ref{eq2}) is satisfied in (d), while not in (e).}\label{fig2}
\end{figure}

Magnetic-field-periodic oscillations are often attributed to the Aharonov-Bohm (AB) effect \cite{PhysRev.115.485,PhysRev.123.1511,PhysRevLett.56.792} .
The AB effect originates from the interference between two wavefunctions propagating along different paths in a ring-like structure (the so-called AB ring).
The AB effect emerges, for example, in the artificially-designed ring-shaped or regularly patterned antidots two-dimensional (2D) devices \cite{PhysicaB.184.398,JPhysSocJpn.81.063707}.
Hence, for the appearance of the AB effect, there must exist an AB ring orthogonal to the magnetic field in the current three-dimensional (3D) sample.
One possible origin is the formation of the moiré superlattice at the stacking faults \cite{JPhysD.38.R329}.
The lattice mismatch arising from the misalignment at the stacking faults leads to the potential variations with a mesoscopic scale,
imposing an interference in the twisted graphene systems \cite{PhysRevB.95.085206}.
In this case, however,
only the interface would be responsible for the conductance oscillations,
which might be masked by the majority part of the bulk.

Another possible origin of the magnetic-field-periodicity is the Sondheimer effect.
The Sondheimer effect has been explained in the semiclassical picture,
and experimentally observed in clean thin films \cite{NatCommun.12.4799,NanoLett.22.65}.
The semiclassical pictures are schematically illustrated in Fig. \ref{fig2}a,b.
As the magnetic field $B$ is applied orthogonal to the conducting thin film with a thickness $d$,
the charged carrier (red ball) moves along the helical orbit (blue solid curve),
which is composed of an out-of-plane free motion and an in-plane cyclotron motion.
The out-of-plane motion with a velocity of $v_{F,z}$ is not perturbed by the magnetic fields,
whereas the in-plane cyclotron motion is determined by the strength of the magnetic field $B$ through the angular frequency of $\omega_{c}=eB/m_{\mathrm{cyc}}$, where $e$ is the elementary charge and $m_{\mathrm{cyc}}$ is the effective cyclotron mass.
As a result, the helical orbit has the same periodicity as the cyclotron motion with a time period $T=2\pi/\omega_{c}$.
Provided that the system is free from any scattering in the bulk,
the carrier travels from the bottom to the top (or \textit{vice versa}) over a time span of $d/v_{F,z}$.
At some special magnetic fields (Fig. \ref{fig2}a),
this time span becomes equal to the integer multiple of $T$;
$d/v_{F,z}=nT$,
where $n$ is an integer.
We refer to this special condition as the Sondheimer condition.
This Sondheimer condition is periodically satisfied with a period $\Delta B$,
which is expressed as \cite{JETP.8.464,PhysRev.172.718}
\begin{equation}
\Delta B = \frac{2\pi}{ed} m_{\mathrm{cyc}} v_{F,z} \propto 1/d. \label{eq1}
\end{equation}
It is noteworthy that the period $\Delta B$ of the Sondhemier effect is inversely proportional to the film thickness $d$.
The Sondheimer effect is in stark contrast to the AB effect for its bulk nature.
In other words, the Sondheimer is innately a 3D feature since it  arises from the out-of-plane motion.
However, the validity of this picture seems limited only in the semiclassical regime,
namely, a substantial number of the LLs participate in the formation of the wave packets
representing the charged carrier in helical motion (red ball in Fig. \ref{fig2}a).
In fact, the Sondheimer effect was treated in the framework of the Boltzmann's theory \cite{JETP.8.464,PhysRev.172.718}.

The magnetic-field periodic oscillations in our graphite thin-film appear in the quasiquantum limit,
where only two LLs with the indices of $N=-1$ and $0$ remain on the Fermi level.
The labels of these indices are in accordance with the convention.
Here, the semiclassical picture is no longer applicable since we cannot compose the particle nature only from the two LLs.
After translating the semiclassical Sondheimer effect formula into the quantum one,
we obtain the condition of the quantum Sondhemier effect as follows;
\begin{equation}
\hbar \omega_{c}=\hbar v_{F,z}(n\Delta k_{z}). \label{eq2}
\end{equation}
Here, $\hbar=h/2\pi$ is the Planck constant divided by $2\pi$, $k_{z}$ is the wave number along the direction perpendicular to the plane, $v_{F,z}=(1/\hbar)\mathrm{d}\varepsilon /\mathrm{d}k_{z}$ is the Fermi velocity along the $k_{z}$, $\varepsilon$ is the energy band dispersion, $\Delta k_{z}=2\pi/d$ is the interval of the $k_{z}$ points, and $n$ is an integer.
Note that $\Delta k_{z}$ is a considerable magnitude for around 100-nm-thick graphite \cite{PhysRevB.97.115122,PhysRevB.98.155136}.
We emphasize that the Eq. (\ref{eq2}) is mathematically equivalent to the semiclassical formula Eq. (\ref{eq1}),
but is expressed in the quantum formalism.
This reformulation leads to the quantum interpretation of the Sondheimer effect at the quasiquantum limit, as shown in Fig. \ref{fig2}c,d,e.
We will only consider $N=-1$ and $0$ LLs at the quasiquantum limit in graphite as a special case
(see Methods for the generalized case).
There are two characteristic energy scales in this situation; the cyclotron energy $\hbar \omega_{c}$ and the discretized energy separation $\hbar v_{F,z} \Delta k_{z}$ (Fig. \ref{fig2}d).
The energy separation between the LLs is characterized by the cyclotron energy $\hbar \omega_{c}$,
which is proportional to the magnetic fields.
Each LL is evenly discretized along the $k_{z}$ direction with an interval of $\Delta k_{z}$ owing to the quantum size effect,
which is solely determined by the thickness and not affected by the magnetic fields.
This discretization corresponds to a standing wave condition along the $z$ direction.
The limitation on the allowed $k_{z}$ mode leads to the discretized energy levels around the Fermi energy $\varepsilon_{F}$ with an energy gap of $\Delta k_{z}\left. (\mathrm{d}\varepsilon/\mathrm{d}k_{z})\right |_{\mathrm{\varepsilon=\varepsilon_{F}}}=\hbar v_{F,z}\Delta k_{z}$.
This separation of the discretized energy levels becomes sizable only in thin film having largely dispersive bands.
With increasing magnetic field,
only the upper $N=0$ LL shifts upwards, whereas the $N=-1$ stays behind \cite{JPhysSocJpn.40.761}.
A point in the discretized level belonging to $N=0$ LL is horizontally aligned to that in the other $N=-1$ LL around the Fermi level when the condition of Eq. (\ref{eq2}) is satisfied,
namely the cyclotron energy is equal to the integer multiple of the discretized energy level separation.
This overlap of the discretized energy levels leads to the resonant inter-LL scattering conserving the energy and the spin of the carriers.
Consequently, the semiclassical Sondheimer effect can be reinterpreted as the quantum Sondheimer effect at the quasiquantum limit.

\begin{figure}[h]%
\centering
\includegraphics[width=0.9\textwidth]{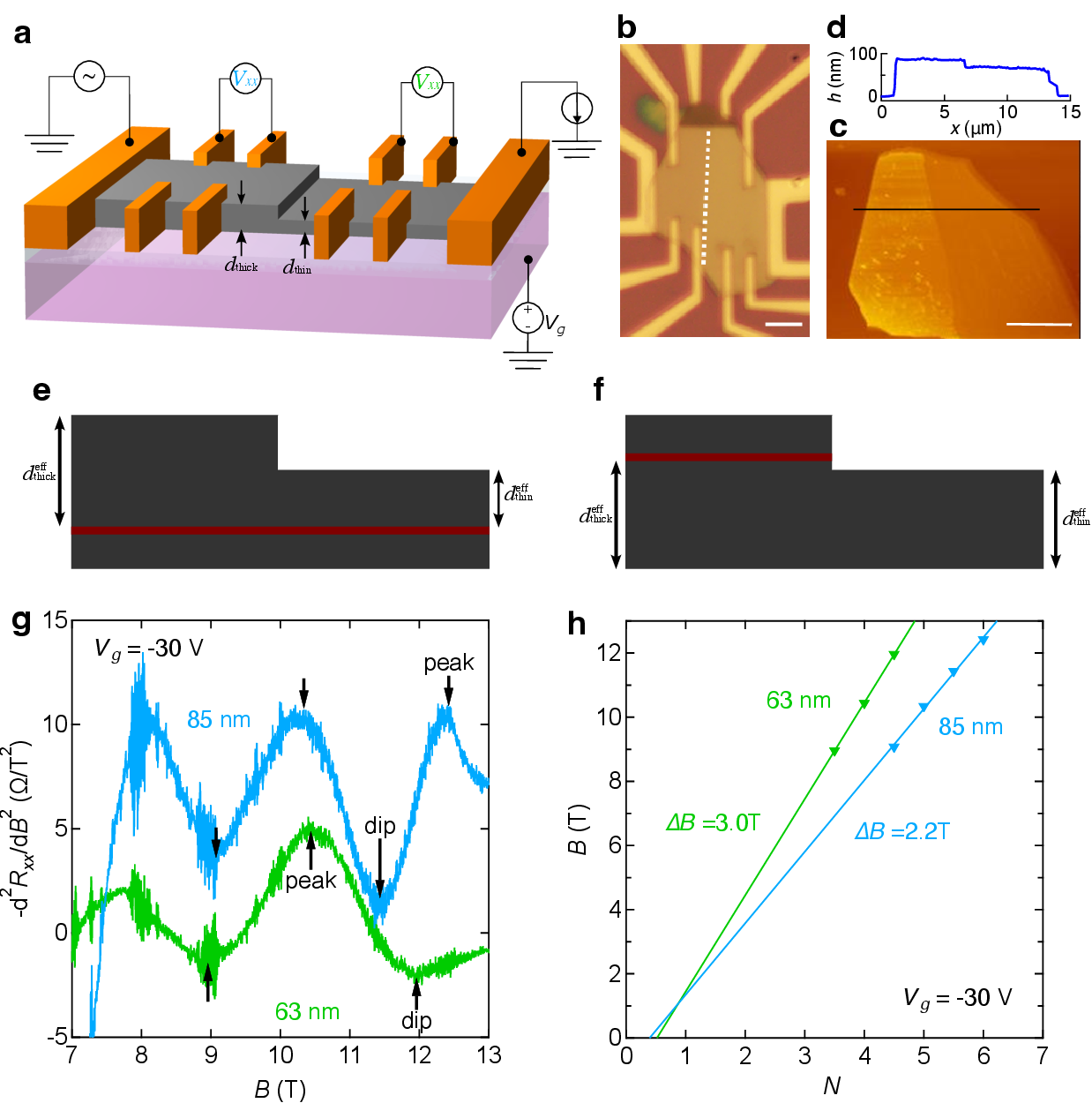}
\caption{Thickness dependence of the $B$-periodic oscillations. (a) Schematic setup of the device with a step in the middle. The thicknesses of the left and right part is $d_{\mathrm{thick}}$ and  $d_{\mathrm{thin}}$, respectively. (b,c) The optical micrography and atomic-force micrography image of the device, respectively. The step is highlighted in a white broken line in Fig. \ref{fig3}(b). The scale bars, 5$\mu\mathrm{m}$. (d) The line profile along the black line in \ref{fig3}(c). (e,f) The schematic side view of the sample with a stacking fault (red line). If the system is scattering free except at the stacking fault, the effective thicknesses of the thick and thin parts are reduced to $d_{\textrm{thick}}^{\textrm{eff}}$ and  $d_{\textrm{thin}}^{\textrm{eff}}$, respectively. (g) The magnetic-field periodic oscillations for the thin (light green) and thick (light blue) parts for $B=7-13\;\mathrm{T}$. The oscillatory components is extracted by the second derivative $\mathrm{d}^2R_{xx}/\mathrm{d}B^{2}$. The bottom gate voltage $V_{g} = -30\;\mathrm{V}$ is applied. Other $V_{g}$ data give almost the same results. The peak and dip positions are marked by arrows. (h) The estimation of the period of the oscillations. By assigning the peak and dip magnetic-field values to the integers and half integers, respectively, the periods $\Delta B$ of the magnetic-field periodic oscillations for thin and thick parts are estimated from the slope of the linear fit as $3.0\;\mathrm{T}$ and $2.2\;\mathrm{T}$, respectively.}\label{fig3}
\end{figure}

The direct evidence of the quantum Sondheimer effect is the relation $\Delta B \propto 1/d$ at the quasiquantum limit.
In order to confirm this relation,
the most simple method is to prepare several samples having various thicknesses
and explore the oscillation periods.
However, the results obtained from different samples need to be carefully compared,
since they might include unintentional differences, for example, the amount of disorder.
In the present case,
stacking faults would play a crucial role,
since those defects obstruct the helical motion.
Although the layered structure of graphite is beneficial to obtain a well-defined thickness,
it is prone to include the stacking faults \cite{JPhysD.38.R329} as a drawback.
The stacking faults segmentize the thickness and break the periodic condition along the out-of-plane direction,
which inhibits the formation of standing waves over the full thickness of the film.
This implies a possible dissociation between the real thickness of the film and the \textit{effective} thickness responsible for the observed oscillations.

In order to avoid ambiguity in the effective thickness,
we designed the sample that enables us to investigate the thickness dependence in a single sample,
as shown in Fig. \ref{fig3}a,b.
The sample has a step in the middle,
which is accidentally formed during the mechanical exfoliation process.
The step divides the sample into the thinner region (right part) and the thicker region (left part)
with a thickness of $d_{\textrm{thin}}$ and $d_{\textrm{thick}}$, respectively.
The thicknesses of the thinner and thicker region in our device were determined from the atomic-force microscopy (AFM) as shown in Fig. \ref{fig3}c,d as $d=63\;\mathrm{nm}$ and $d=85\;\mathrm{nm}$, respectively.
It is natural to assume that the stacking faults, such as twisted interface, were formed at the crystal growth,
so the microcrystal for the device, which is exfoliated from the as-grown crystal, includes those stacking faults over its whole region (Fig. \ref{fig3}e).
On the other hand, in the mechanical exfoliation process,
the edge parts of the microcrystal are subject to the force to peel off layers,
which could introduce planar cracks that are concentrated at the edge.
Most of these planar cracks are expected to disappear
since the two layers across the planar cracks can combine via the van der Waals interaction
as long as the gap of the cracks is small enough not to be wedged by obstacles.
A small number of remaining cracks
might have no effect on the transport properties since they exist only at the sample edges.

If the sample includes a stacking fault over its entire region,
as shown in Fig. \ref{fig3}e,
both the thinner and thicker regions would have reduced effective thicknesses,
denoted as $d_{\textrm{thin}}^{\textrm{eff}} (<d_{\textrm{thin}})$ and $d_{\textrm{thick}}^{\textrm{eff}} (<d_{\textrm{thick}})$, respectively.
In the case of Fig. \ref{fig3}e,
even if we cannot determine the absolute values of these effective thicknesses,
the trend of the thickness-dependent transport features can be detected
since the thickness relation holds ($d_{\textrm{thin}}^{\textrm{eff}} < d_{\textrm{thick}}^{\textrm{eff}}$).
Another possible configuration is shown in Fig. \ref{fig3}f, where a stacking fault only exists in the bump region.
Even in this case,
although the thickness difference between thinner and thicker parts becomes small or zero,
the relation is not inverted ($d_{\textrm{thin}}^{\textrm{eff}} \leq d_{\textrm{thick}}^{\textrm{eff}}$).
As a result,
the trend of thickness dependence of the transport property can be observed by using this step-structure sample even if the stacking faults are included.

The observed resistance oscillations ($\mathrm{d}^2R_{xx}/\mathrm{d}B^{2}$) for the thinner (green curve) and the thicker regions (light blue curve) are plotted in Fig. \ref{fig3}g.
As expected from the quantum Sondheimer effect ($\Delta B \propto 1/d$, Eq. (\ref{eq1})),
a larger period is identified in the thinner part of the sample.
By assigning the peaks and the dips to integers and half integers, respectively,
the period of each region is determined, as shown in Fig. \ref{fig3}h.
The slope of the fitting line yields the period of the magnetic-field-periodic oscillations,
with the value of $\Delta B = 3.0\;\mathrm{T}$ and $2.2\;\mathrm{T}$ for $d = 63\;\mathrm{nm}$ and $85\;\mathrm{nm}$ regions, respectively.
The ratio of the periods $3.0/2.2\approx 1.36$ is in good agreement with the inverse-thickness ratio $85/63=1.35$,
which is consistent with the quantum Sondheimer effect.
Note that this quantitative agreement suggests that the stacking faults are absent in this device.

The relation between the period $\Delta B$ and the inverse of the thickness $1/d$ is summarized in Fig. \ref{fig4}a (orange diamonds).
The periods obtained from step-free samples \cite{PhysRevB.98.155136} are also displayed (black solid circles).
Regardless of the sample qualities,
almost all the data points fall onto the line determined by the sample with a step,
which is presumably free from the stacking faults.
The consistent linear relation between $\Delta B$ and $1/d$ indicates that the magnetic-field-periodic oscillations in the quasiquantum limit are attributable to the quantum Sondheimer effect.
Although multiple periods with exceptionally large values are found in the sample with $d=154\;\mathrm{nm}$ ($1/d=0.65\times 10^{-2}\;\mathrm{nm}^{-1}$),
the result is reasonably explained by the segmentation of the thickness by the stacking faults.
If we assume that each segmented part produces the oscillations,
the observed periods are well explained by the same line in Fig. \ref{fig4}a,
supporting the quantum Sondheimer effect scenario.

\section*{Discussion}\label{sec3}
\begin{figure}[h]%
\centering
\includegraphics[width=0.9\textwidth]{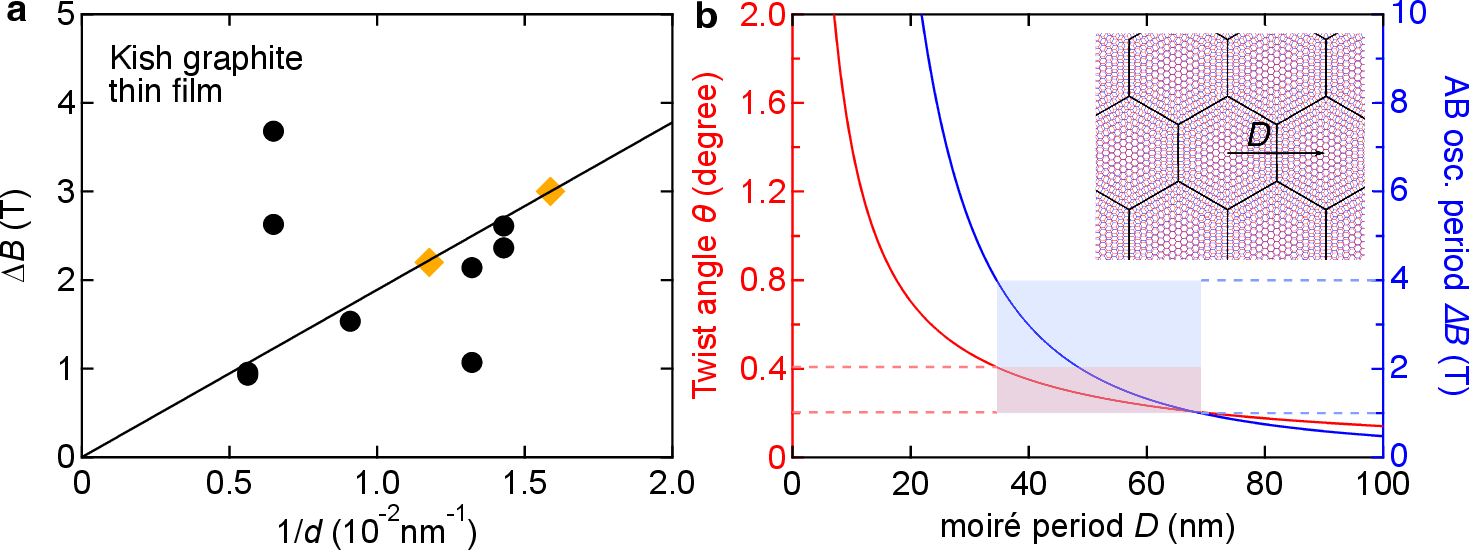}
\caption{The relation between magnetic-field period and the characteristic length scale. (a) Experimentally observed magnetic-field periods as a function of the inverse of the thickness $1/d$. Orange diamond markers are from the device with a step. The solid line is the guide to the eye. (b) The consideration of the AB effect scenario. The moiré period $D$ (bottom axis) is determined by the twist angle $\theta$ (left axis) through the formula $D=a/(2\sin (\theta/2))$ (red line). By assuming the area of the moiré unit cell $S$ enclosed by an AB ring, the AB oscillation period $\Delta B$ (right axis) is expressed as $\Delta B =\phi_{0}/S$ (blue line). In order to assign the experimentally observed oscillations to the AB oscillations, the moiré period should be $35-69\;\mathrm{nm}$, which corresponds to the limited range of the twist angle $\theta\approx 0.2-0.4\;^{\circ}$. The inset indicates the moiré superlattice formed at the stacking fault with a moiré period $D$.}\label{fig4}
\end{figure}

Our results indicate that the AB effect originating from the superlattice at the stacking-fault interface is unlikely to emerge.
First, only a very narrow range of the twist angle $\theta$ is possible to explain the observed oscillations period.
The moiré period $D$ (the period of the so-called AA stacking region, inset of Fig. \ref{fig4}b) formed at the stacking fault of graphite is determined solely by the twist angle through the equation $D=a/(2\sin (\theta/2))$ (red curve in Fig. \ref{fig4}b),
where $a$ is the in-plane lattice constant of the graphite.
On the other hand,
the AB oscillations period $\Delta B$ is determined by the area $S$ enclosed by the AB ring through the equation $\Delta B =\phi_{0}/S$,
where $\phi_{0}=h/e$ is the magnetic flux quantum.
By assuming that each moiré unit cell (hexagonal cell enclosed by black solid lines in the inset of Fig. \ref{fig4}b) works as a unit of an AB ring,
$S$ can be represented with moiré period $D$ through the equation $S=\sqrt{3}D^2/2$,
namely, the AB oscillations period $\Delta B$ is determined by the moiré period $D$ (blue curve in Fig. \ref{fig4}b).
If the observed magnetic-field-periodic oscillations with a period of $\Delta B = 1-4\;\mathrm{T}$ are attributed to the AB effect,
the value of the moiré period should be $D \approx 35-69\;\mathrm{nm}$ (blue shade in Fig. \ref{fig4}b).
This moiré period can be translated into the twist angle at the stacking fault as $\theta \approx 0.2^{\circ}-0.4^{\circ}$ (red shade in Fig. \ref{fig4}b).
Since the twist angle at the stacking fault should have a random value,
it is very unnatural to assume that only a very limited range of twist angles are formed.
In other words,
random $\Delta B$ values should be observed if it was due to the AB effect mechanism.
Although the AB-ring area $S$ would be quantitatively different from the current estimation when the crystal reconstruction at the interface occurs,
it is difficult to reproduce such a limited range of oscillation periods by using the AB effect scenario.

Finally, we summarize the necessary condition to observe the quantum Sondheimer effect.
First, a clean bulk with a well-defined thickness is necessary.
This condition was realized in a clean semimetal fabricated using a focused-ion beam method \cite{NatCommun.12.4799},
but it is easily satisfied in layered materials.
Second, the energy spacing arising from the quantum size effect $\hbar v_{F,z} \Delta k_{z}$ needs to be resolved.
This condition requires a large Fermi velocity $v_{F,z}=(1/\hbar)\mathrm{d}\varepsilon /\mathrm{d}k_{z}$,
i.e., a large bandwidth.
Third, only a few LLs should remain at the Fermi level at the accessible magnetic fields in order to reach the \textit{quantum} Sondheimer regime.
In the graphite case, only two LLs remain above $B=7.4\;\mathrm{T}$ neglecting the spin degrees of freedom.
As a result, graphite thin film turns out to be an ideal platform to satisfy these conditions.
The quantum Sondheimer effect should be generically observed as long as these conditions are satisfied.
The combination with other electronic states, such as the density-wave state, could offer new paths to explore the novel physics in quantum devices.

\section*{Methods}
\subsection*{Sample fabrication}
Our thin-film graphite was prepared by mechanically exfoliating Kish graphite crystals using Scotch-tape, followed by transfer onto Si/SiO$_{2}$ substrate,
electron-beam lithography, and a deposition of Cr/Au contact.
A heavily-doped silicon substrate was used for applying gate voltage $V_{g}$,
which did not play a important role in the present study.
A typical sample dimension is around $50\times 50\;\mu\mathrm{m}$ in the plane and a thickness is of the order of $100\;\mathrm{nm}$.
The thickness is over hundred times larger than the out-of-plane lattice constant,
which ensures that the system is far beyond the two-dimensional limit.
Each sample thickness was determined by AFM analyses.
A sample with a step in the middle, as shown in Fig. \ref{fig3}(a)-(c),
was identified by using polarized optical microscopy,
confirmed by AFM observations.

\subsection*{Transport measurements}
The in-plane electrical resistance $R$ was measured by dc method with reversing a constant current under a magnetic field $B$ along the $c$ axis.
High magnetic fields were generated in a 35 T resistive magnet at the National High Magnetic Field Laboratory.
The sample was cooled down to 0.35 K by using helium-3 refrigerator.
With regard to the sample with a step,
the measurement configuration was the same as above.
The resistance was measured with lock-in technique at the temperature of 4 K and the maximum magnetic field of 13 T,
which is generated by a superconducting magnet system.
The exhibited $R_{xx}$ in the step sample was symmetrized with respect to the magnetic field.

\subsection*{Spin degrees of freedom}
We discuss the quantum Sondheimer effect in graphite by neglecting the spin degrees of freedom.
In the realistic band structure of graphite,
there are four spin-split Landau bands at the quasiquantum limit,
$(N=0, \uparrow), (N=0, \downarrow), (N=-1, \uparrow)$, and $(N=-1, \downarrow)$.
However, this spin splitting is not important
since the resonance condition requires the coupling between the energy and spin conserved states.
The spin degrees of freedom can be taken into account
by considering two distinctive pair of $(N=0, \uparrow) \leftrightarrow (N=-1, \uparrow)$ and $(N=0, \downarrow) \leftrightarrow (N=-1, \downarrow)$.

\subsection*{General case of the quantum Sondheimer effect}
\begin{figure}[h]%
\centering
\includegraphics[width=0.9\textwidth]{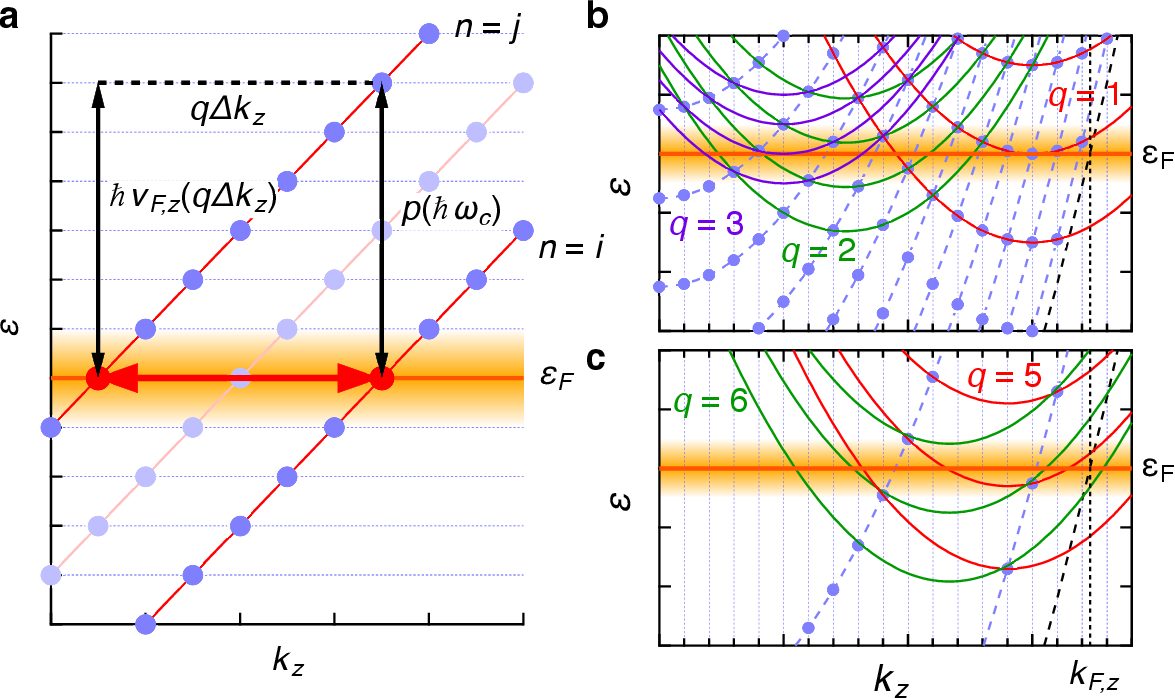}
\caption{The general case of the quantum Sondheimer effect. (a) The resonant pair (red markers) on the Fermi enrgy $\varepsilon_{F}$ with $p=j-i$ and $q$ associated with the quantum Sondheimer effect. The band dispersions are approximated with a linear one. 
(b,c) The Sondheimer subbands (solid lines) under lower (b) and higher (c) magnetic fields on the basis of Landau subbands for the free electron model (Eq. (\ref{eq4}), broken blue lines). Only the Sondheimer subbands with $p=1$ are shown.
}\label{fig5}
\end{figure}

At the quasiquantum limit in graphite, only two LLs of $N=0$ and $N=-1$ remain at the Fermi energy.
Moreover, since the $N=-1$ LL is insensitive to the magnetic field,
it is easy to derive the Sondheimer condition Eq. (\ref{eq2}).
Here we will consider the general case,
where an arbitrary number of LLs exist with an equal energy spacing $\hbar \omega_{c}$.
The key concept of the quantum Sondheimer effect arises from the resonant states among the discretized energy levels belonging to different Landau indices.
Suppose the resonant pair is found between the LLs indexed by $N=i$ and $N=j=i+p$ (Fig. \ref{fig5}a),
where the integer $p=j-i\geq 1$ indicates the Landau index difference.
At one $k_{z}$, the energy difference between the discretized levels belonging to $N=i$ and that to $N=j$ is equal to $p(\hbar \omega_{c})$.
In addition,
another integer $q\geq 1$ is introduced so as to minimize the energy difference $p(\hbar \omega_{c})-\hbar v_{F,z} q\Delta k_{z}$,
where the Fermi velocity along $k_{z}$ direction $v_{F,z}$ gives the slope of the Landau subband through $\hbar v_{F,z}$.
If this energy difference goes zero,
a resonant pair is formed (red markers in Fig. \ref{fig5}a).
As a result,
the general formulation of the quantum Sondheimer effect is described as
\begin{equation}
	\frac{p}{q} \hbar \omega_{c}=\hbar v_{F,z} \Delta k_{z}.
	\label{eq3}
\end{equation}
For each Landau-level configuration,
multiple resonant modes indexed by a ratio $p/q > 0$ are allowed.
The quasiquantum limit is the special case,
where at most one resonant mode $(p/q=1/n)$ survives at a fixed magnetic field (Eq. (\ref{eq2})).

In order to unify the quantum and the semiclassical pictures,
the fully-quantum calculation of the conductivity is required in the semiclassical regime,
which is beyond the scope of this paper.
Instead, hereafter we will see a trace of the semiclassical characteristics by using the quantum picture.
In the semiclassical Sondheimer picture, illustrated in Fig. \ref{fig2}a,b,
the dominating contributions for the oscillatory part of the conductivity $\Delta \sigma$ arise from the local Gaussian curvature of the Fermi surface at the limiting points (extreme $k_{z}$) for the elliptic Fermi surface \cite{JETP.8.464,PhysRev.172.718,PhysRevB.8.5567}.
The Landau quantization and the concomitant quantum Sondheimer resonance are considered in the situation where a substantial Landau levels exists.
For simplicity, we focus on the free electron model,
where each LL dispersion is written as
\begin{equation}
	\varepsilon_{N} (k_{z}) = \left(N+\frac{1}{2}\right)\hbar \omega_{c} + \frac{\hbar ^{2} k_{z}^{2}}{2m}.
	\label{eq4}
\end{equation}
By introducing the quantum size effect,
$k_{z}$ is discretized with a unit of $\Delta k_{z}=(2\pi)/d$.
The discretized Landau subbands for lower and higher magnetic fields are shown in Fig. \ref{fig5}b,c, respectively (blue broken lines are the guide of Eq. (\ref{eq4})).
The conventional quantum oscillations arise from the bottom of these Landau subbands passing through the Fermi energy $\varepsilon_{F}$.
In addition to the Landau subbands,
another subbands (referred to as the Sondheimer subbands), exemplified by solid lines, are visualized in Fig. \ref{fig5}b,c.
For a substantial number of LLs,
a series of $p/q$ Sondheimer subbands satisfies the Sondheimer condition Eq. (\ref{eq3}).
The analytical expression of the solid lines in Fig. \ref{fig5}b,c is
\begin{equation}
	\varepsilon_{\{N_{0},j\}}^{(p/q)} (k_{z})
	=\left( N_{0} + \frac{1}{2}+j\frac{p}{q}\right) \hbar \omega_{c} 
	+\frac{\hbar ^{2}}{2m_{z}} \left(k_{z}-k_{z,\mathrm{min}}^{(p/q)}\right)^{2}
	-\frac{\hbar ^{2}}{2m_{z}} \left(k_{z,\mathrm{min}}^{(p/q)}\right)^{2},
	\label{eq5}
\end{equation}
where $j$ is an integer such that $0 \leq j \leq q-1$,
$m_{z}$ is the effective mass along $z$ direction,
$k_{z,\mathrm{min}}^{(p/q)} = \frac{p}{q}\frac{deB}{2\pi \hbar} = \frac{p}{q}\frac{1}{(\hbar^{2}/2m_{z})\Delta k_{z}}\frac{\hbar \omega_{c}}{2}$ is the wave number giving the Sondheimer subband minimum,
and $\{N_{0},j\}$ is the Sondheimer subband indices.
It is noteworthy that the minimum location $k_{z,\mathrm{min}}^{(p/q)}$ for each $p/q$ mode linearly shifts towards higher $k_{z}$ by sweeping up the magnetic fields (Fig. \ref{fig5}c).
If both two consecutive points in one $p/q$ subband locate together around the Fermi energy,
that $p/q$ mode satisfies the Sondheimer condition Eq. (\ref{eq3}).
Therefore, only the $p/q$ modes satisfying $k_{z,\mathrm{min}}^{(p/q)} < k_{\mathrm{F},z}$ are relevant.
Among these modes, the largest $p/q$ mode, which gives the largest $k_{z,\mathrm{min}}^{(p/q)}$, plays a main role,
since it is the first one to be excluded from them by increasing the magnetic field.
Every time the largest $p/q$ mode goes beyond the limiting point, namely $k_{z,\mathrm{min}}^{(p/q)} \gtrsim k_{\mathrm{F},z}$,
the contribution of that $p/q$ mode is eliminated from the sum of the resonant modes.
This quantum picture emphasizes the importance of the behaviour around the limiting point,
which corresponds to the semiclassical picture.

\bmhead{Acknowledgments}
The authors acknowledge Dr. E S. Choi for technical support and discussions. A portion of this work was performed at the National High Magnetic Field Laboratory, which is supported by National Science Foundation Cooperative Agreement No. DMR-1157490 and the state of Florida. This work was partially supported by JSPS KAKENHI Grants No. JP15K21722, No. JP25107003, No. JP16H03999, No. JP16K17739, JP19K14655, and JP22K03523.

\subsection*{Author Contributions}
T.T. conceived and designed the experiment.
T.T. fabricated samples, performed transport measurements, and analyzed data.
T.O. performed the theoretical analysis, with help from A.K. and T.T.
All authors participated in discussions and in writing of manuscript.

\subsection*{Declarations}

The authors declare no competing interests.



\end{document}